\documentclass[preprint]{rmaa}
      
\title{Catalog of  156 Confirmed Extrasolar Planets and Their
133 Parent Stars}
\author{J. Espresate \altaffilmark{1}}

\altaffiltext{1}{Instituto de Astronom\'{\i}a, UNAM, Ciudad
  Universitaria, M\'exico D.F., M\'exico} 

\fulladdresses{
\item J. Espresate: Instituto de Astronom\'{\i}a, UNAM,,
  Apartado Postal 70-264, 04510 M\'exico, D.F., M\'exico
  (julia@astroscu.unam.mx)}
    
\shortauthor{Espresate} \shorttitle{Extrasolar Planets Catalog}

\ReceivedDate{} \AcceptedDate{} \SetYear{2005}
\resumen{En este art\'{\i}culo presento dos cat\'alogos, el primero
contiene
los datos observados e inferidos 
de las estrellas que poseen planetas extrasolares 
confirmados. Los datos que reporto para estas 133 estellas son:
masa estelar; tipo espectral;
clase de luminosidad; periodo de rotaci\'on; 
metalicidad [Fe/H]; luminosidad absoluta;
radio estelar;
temperatura estelar; edad y n\'umero de planetas que posee.
El segundo cat\'alogo, 
contiene los datos de
los 156 planetas extrasolares que giran alrededor de las 133 estrellas.
Este cat\'alogo re\'une todos los
datos disponibles de los planetas extrasolares: masa m\'{\i}nima y
elementos orbitales; inclinaciones; radio del planeta; distancia
promedio a la estrella; signo de la torca
de marea (si se conoce el periodo de rotaci\'on estelar); la
Irradiancia sobre el planeta suponiendo un albedo geom\'etrico de 0.
La ventaja
de este trabajo es reunir el mayor n\'umero de datos observados o
inferidos tanto de los planetas extrasolares como de la estrellas a
las que orbitan, en un solo cat\'alogo. 
Se reportan varios planetas cuya \'orbita
est\'a {\sl totalmente contenida dentro de la \'orbita sincr\'onica 
de la estrella.}}

\abstract{I present two catalogs. The first one contains
 the observed and infered data of all of 133
stars wich harbor confirmed extrasolar planets. 
For these 133 stars I report
the following data: stellar mass; spectral type; luminosity class;
stellar rotation period;
stellar metalicity, [Fe/H]; absolute luminosity; stellar radius; 
stellar temperature; age and number of planets host by the star.
The second catalogue lists data for 156 confirmed 
extrasolar planets orbiting around the 133 stars.
This catalogue contains the following extrasolar planet data:
minimum mass and orbital elements; inclinations; planet radius; 
average distance to its parent star; 
tidal torque sign (if the
stellar rotation period is reported); total irradiance 
assuming zero albedo.  
The usefulness
of these catalogs is that they contain in a single paper all of the
available data of the extrasolar planets, 
and their parent stars. Several planets whose orbits are {\sl completely
contained inside the star's syncronous orbit, are reported.}}

\begin{document}
\maketitle
\tableofcontents

\section{Introduction}
\label{sec:intro}
In the last  decade 133 stars have been reported to harbor
planets, however the total number of confirmed extrasolar
planets orbiting these stars is 156, 
implying that
some stars harbor more than one planet. 
So far the number of multiple exoplanets systems
is 17 (Schneider, 2005).

The vast majority of the planets have been discovered through
spectroscopic thechniques from which the periodic variation of the
star's radial velocity (component in our line of sight) is detected. 
This periodic motion results from perturbations on the star's motion
due to one or more planetary 
companions. Since the inclination of 
the orbit with respect to the plane of the sky is not known,
spectroscopic measurements gives us only a minimum for the star's 
velocity around the common 
center of mass and hence a minimum mass for the perturber
or perturbers.

There is a strong bias towards large planets with small semimajor
axes, because these are the ones that produce the largest perturbations
on their parent star motion. Small planets produce very small 
perturbations even if they orbit close to the star and may be 
undetectable with the 
current precision.
Another interesting bias
is for large planets on faraway orbits due to their long orbital
periods. In this  case the star has to be 
observed for several years in order to obtain a
good  radial velocity curve that later can be fitted assuming
the gravitational interaction with one (or more) 
large planetary companion.

The unknown
quantity is always the inclination of the orbit with respect to the
plane of the sky 
except for the few so called ``transiting planets''. 
The larger the inclination is
(towards 90 degrees), the more similar is the measured star's 
radial velocity to its
total velocity and hence the more accurate is
the inferred mass of the planet. When the inclination is unknown, the
fitting of the radial velocity curve
provides a minimum mass for the planet.
 
Two observing groups have made most of the extrasolar planets
discoveries.
The Anglo-Australian group, lead
by Geoeffrey W. Marcy and R. Paul Buttler  
known as the California \& Carnegie Planet Search Team
which obtain
their measurements using several telescopes 
mainly, the Keck Observatory,
the Lick Observatory and Mac Donald's observatory. The other
group is the Geneva Extrasolar Planet Search Programmes
lead by M. Mayor, D. Naef,
F.Pepe, D. Queloz, N.C. Santos and S. Udry.
Their program includes
the Coralie Survey for Southern Extrasolar Planets (La Silla, Chile),
the ELODIE Northern Extrasolar Planet Search (OHP-France) and the
M-Dwarf Pogrammes.
Users of this 
catalog are welcome to visit the internet pages and/or papers(see
references) to find out more details about the
instruments and detection methods of the various
quantities reported here.
Most of the very basic information about 
the extrasolar planets can be
found in the Internet without much effort.
However, {\it when it comes to the
characteristics} of the parent stars, things become really entangled 
because the information is spreaded in numerous papers and catalogues,
over many years.

All I did was to gather specially the hosts stars data  and organize
them in a useful form so they can be found immediatelly. In some cases
there are non-negligible differences for the same quantity as reported
in different papers or lists. For instance, the spectral
type is sometimes controversial or simply not determined.
Minor variations that my be considered negligible or not,
(depending on the calculation one is trying to make), are
the rotation period of the star, the planet's orbital excentricity
and or minimum mass. On
each case I decided to take the data as reported by the discoverers
either on the web or on the paper itself. 
If a star is classified between
two spectral types and one of them is main squence,
then I classified it arbitrarily as a main sequence star. 
When possible, 
incomplete spectral types reported by discoverers 
were taken from Simbad if available.
Other non-reported quantities are calculated here.

\section{133 Host stars} 
\label{sec:star catalog}

The first table presents the stars in essentially the same order as
found in Schneider
Catalog (2005) hereafter (SC). 
That is, by increasing semimajor axis of the
exoplanets. The star's identification
is by their Henry Draper (HD) number
unless they do not have one. In this case the identification is the
number and letters of the
most known catalog in which they appear. No constelation related names
are used. 

I wanted a list as complete as possible 
for the host stars data, unfortunately for several of them
I could not find either, 
effective temperatures (T$_{eff}$),
Luminosities, 
(L$_*$), radius (R$_*$), or any of the aforementioned quantities.
Therefore I decided to proceed as follows:
I divided the whole set of 133 stars in  6 groups, 
depending on which of the three
aforementioned quantities was missing 
in the reported information:

{\it Case 1}

If none of the three quantities were found {\it and the star 
is classsified as main sequence}, I calculated them 
through the Kippenhan models using first
the mass-radius relation as follows:

$${R_*\over R_\odot} = \Bigg({M_* \over M_\odot}\Bigg)^{\eta}
 \eqno (1)$$

\noindent
where $\eta =0.8$ if $M_* \leq 1 M_\odot$ and $\eta =0.57$ otherwise,
which are the average exponents I obtained from Kippenhan's 
plots of log[$M_*/M_\odot$]
vs. log[R$_*$/R$_\odot$]. Then the luminosity is calculated
through  the 
mass-luminosity relationship between $-1 \le$log[ M$_*$/M$_\odot$]
$\le 1.5$
which gives an exponent $\beta = 3.2$, that is:

$${L_*\over L_\odot} = \Bigg({M_* \over M_\odot}\Bigg)^{\beta} 
\eqno (2)$$

\noindent
and from there the effective temperature considering the star
as a perfect 
black-body:

$$T_{eff}= \Bigg({L_* \over {4\pi R_*^2\sigma}}\Bigg)^{1/4}. \eqno(3)$$

\noindent
The total number of stars in this case is 12.

{\it Case 2}

The star's radius is reported but no temperature or luminosity. 
In this case
I kept the reported radius, calculated L$_*$ using Eq.(2), and obtained
T$_{eff}$
from Eq.(3). Only 2 stars were in this case.

{\it Case 3}

No stars with  reported luminosity but no 
radius and no temperature reported.

{\it Case 4}

Effective temperature reported but no luminosity and no radius. I kept 
the reported value of T$_{eff}$ and calculated the radius with Eq.(1), 
and then obtained
the luminosity solving for L$_*$ in Eq.(3). 
There were 4 stars in this case.

{\it Case 5}

Luminosity and temperature are reported for the star but not
its radius. I kept the values reported and found R$_*$ from Eq.(3).
There are 23 stars in this case.

{\it Case 6}

Temperature and radius are reported, but no Luminosity. 
I solved Eq.(3) to obtain the 
luminosity using the reported values of the other two queantities. 
There are 30 stars
found in this case.

For the first 5 cases
I used the Kippenhan Models which
are exclusively valid for main sequence stars (luminosity class V).
Hence I only applied the Kippenhan models to main sequence stars.
Recall these models do not take into account
the metallicity of the stars.
Therefore the calculations are 
only good approximations to the unfound data.

Depending its case, each star has a superindex right after the 
identifier (or name) indicating which quantities were calculated 
and which were found in the litrature.

In Table 1, I list the following stellar data:

{\it Column (1)} Reference number in this catalog.

{\it Column (2)} Star's identifier.

{\it Column (3)} Spectral type and Luminosity class.

{\it Column (4)} Stellar mass in M$_\odot$.

{\it Column (5)} Luminosity in L$_\odot$.

{\it Column (6)} Effective Temperature in Kelvin.

{\it Column (7)} Metallicity [Fe/H] dex.

{\it Column (8)} Rotation period in days.

{\it Column (9)} Stellar Radius in R$_\odot$. 

{\it Column (10)} Age in Myr

{\it Column (11)} Number of planets host by this star

\vskip0.3cm

For not main sequence stars which had no luminosity reported,  
but found estimated
stellar radius and effective temeperature,
I used Eq. (3) to calculate the luminosity.
All of the stars whose luminosities were calculated in this way,
have a superindex $^+$ right after their identification name.

For the main sequence star No. 73, HD12661 for which I found no data of
its luminosity I took the reported 
radius and effective temperature given in Fischer {\it et al.} (2001)
and used Eq. (3) and solved for Luminosity.

For the star "HD219449" (No. 49) I could not find the
data of its mass, 
which is surprising to me since the orbit of its planet is
apprently well determined except for its excentricity.
Therefore for this star which is a K0III, 
no calculations were performed, 
it has no metallicity reported either, or
rotation period. 

Finally 
I excluded star OGLE-235/MOA-53 which appears in SC, 
because I did not find enough data, not only about 
the planet but also, and specially about the star.

\section{Exoplanets catalog} 
\label{sec:exoplanets}

In Table 2, I report the propperties of 156 
exoplanets sorted by increasing semimajor axis of the planet as 
they appear in SC
catalogue, except that I took the values directly from the literature. 

There are two stars for which the existence of a 
second planet was controversial until very recently. 
The star HD 128311 (No. 86)
appears with one planet in SC and two planets in Marcy's
catalog. The second planet has been already confirmed (Voght, {\it
et al.} ApJ preprint, 2005), therefore I listed the data they report
in this last paper for the two planets around HD 128311.

The system HD 41004 is a close visual binary
composed by a K2V star (HD41004A) plus a M4V star (41004B)
(see http://obswww.unige.ch/~udry/planet/hd41004A.html).
Both stars host one planet each, as reported by the
CORALIE team. I put the information as presented by them;
the companion of HD 41004B (No. 120 in Table 2)
has a minimum mass of 18.37 M$_{J}$
and is interpreted as a brown dwarf (Santos {\it et al.} 2002).
However its host star  
{\bf HD 41004B is not reported in Table 1}, because
I found no data on this star except its spectral type. On the other
hand HD41004A is the star No. 99 in Table 1 and its
planet is No. 119 in Table 2. 

The 156 exoplanets (icluding the brown dwarf mentioned above) 
are listed in the same order as their parent stars  
in Table 1, that is, 
by increasing semimajor axis.
All of their identification  names are the 
name of its host star followed by a letter {\bf b, c, d} etc., 
according to the number of planets that the star hosts. 
The order of the letters 
{\bf b, c, d} is not related to the distance between the planet and 
the star, it only represents the cronological
order of their discovery.

Table 2
 contains the following quantities:

{\it Column 1} Number of planet in this catalog.

{\it Column 2} Name of the planet

{\it Column 3} Mass of the planet in Jupiter masses (M$_J$)

{\it Column 4} Planet radius in Jupiter Radius (R$_J$)

{\it Column 5} Semimajor axis in Astronomical Units (AU)

{\it Column 6} Eccentricity

{\it Column 7} Orbital period (days)

{\it Column 8}  Average orbital radius (AU) 

{\it Column 9}  Irradiance (ergs cm$^{-2}$ s$^{-1}$)

{\it Column 10} Apoapse distance r$_a$ divided by the radius of the
star's syncronous orbit r$_s$.

{\it Column 11} Inclination (degrees)

All of the planet data reported here come from the literature, and 
although some differences appear between different authors, I chose again
the data as reported by the discoverers (see references).

In what follows I explain the calculations made in this work and not
found in the literature.

{\it Column 9} Is the average ammount of energy, $I$, in units of $10^6$
ergs, per 
square centimeter, per second that reaches the planet's 
surface assuming zero albedo, that is:

$$I = {L_*\over 4\pi a^2(1+e^2/2)^2} \ \ \ \ , \eqno(4)$$

\noindent 
where $r_a=(1+e^2/2)$ is the average distance of the planet to the star.

Finally, {\it Column 10} is the planet's orbit apoapse r$_a=a(1+e)$
divided by the radius of the star's syncronous orbit.
The radius of the star's syncronous orbit, $r_s$, is given by:

$$r_s=\Bigg({P_{rot}^2 G M_* \over 4\pi^2}\bigg)^{1/3}\ \ \ , \eqno(5)$$

\noindent where
$P_{rot}$ is the star's rotation period, $M_*$ is its mass, and $G$
is the gravitational constant.
This calculation is only possible (of course) 
if the star has a rotation period reported.

The very basic assumption behind this calculation is that at least, to
a first aproximation,
the orbital angular momentum of the planet is parallel and has the
same direction as the spin
angular momentum of the star. Based on this assumption one can get
from this calculation an approximate idea of
which of these extrasolar planets are
inside or outside 
the syncronous orbit.

If this ratio is
less than 1 then the planet's orbit is completely contained inside the
syncronous
orbit and therefore it is subjected to
a negative torque which decreases its orbital angular 
momentum (and energy). This 
decrease is due to the lag of the star bulge raised by the planet. 
Being inside the synchronous orbit, the planet revolves around the star
faster than the star rotates. Therefore the tidal
bulge  raised on the star by the planet points to a certain angle that falls
behind the planet angular position at any time in its orbit, causing a decrease
in the planet's angular momentum and energy. If no other forces are considered
the unavoidable fate of the planet is to eventually fall onto the star, however
slow. 

If this ratio is grater than
1, then the planet is "safelly" outside the syncronous orbit and the 
average tidal torque is positive. The star's bulge is always pointing ahead
the planet's angular position and therefore,
the planet is increasing its orbital angular momentum and energy  
moving outwards however slow, because the bulge that it raises on the star
is probably very small.

\section{Discusion and results} 
\label{sec:discusion} 

Before I fully enter the discussion of results I present the following few
statistics made over this star and planet sample.
Table 3 shows the number and percentage over thew hole either star or
planet sample of the characteristics indicated in the fist column. 

\linespread{1.1}
\begin{table}[h]
  \setlength{\tabcolsep}{0.1em} 
 \begin{center}
    \caption{statistics}
    \label{tab:3}
    \begin{tabular}{lcc}\hline\hline
Propperty& Number & Percentage \\
\hline
Luminosity class V & 98 & 73.68 \\
Luminosity class III & 5 & 3.76 \\
Luminosity Class IV & 15 &11.28 \\
No Luminosity Class reported& 15&11.27\\
Stellar rotation reported&81&60.9\\
Metallicity Reported&131&98.5\\
Negative Tidal Torque & 22 & 14.1 \\
Planets with $a$ less than 1 AU& 59& 37.8 \\
   \hline\hline
       \end{tabular}
  \end{center}
\end{table}
\linespread{1.6}

From the previous table one can see that there is a serious lack of
measurements of stellar rotation periods. Almost $\sim$ 40 \% of the
star sample does not have a reported rotation period. On the other hand
almost all of the stars have a measurement of $[Fe/H]$; however, I must warn
the reader that these reported measurements come from different works, and
are not necessarilly consistent amongst them in the sense of callibrations, 
instruments and or methods. Nevertheless these values are useful to make 
statistics or even approximate theoretical models.

About 10\% of the star sample have no determined Luminosity Class which is
an important parameter  for studies of extrasolar planets around evolved stars.

To me, the most striking result in this paper 
is the 22 planets whose orbits are 
completely contained inside the synchronous orbit of the star.

All migration models assume  that planets formed relatively far away
from the star and migrate inwards. Aside from the main controversies 
around this process in its several scenarios, they all begin with a flat
disk around the star through which the formation and migration of 
the planets take place. The newly formed planets loose angular 
momentum by interactions with the inner and outer parts of 
this keplerian disk. Hence it is consistent to assume that most probably
the majority of planets formed in these disks, have prograde orbits and 
therefore as in our own Solar System their {\it orbital angular momenta}
are parallel and in the same direction as the spin angular momentum of their
stars. These 22 planets pose a new question:
if these planets arrived to their present 
positions through migration, how did they manage to cross the 
synchronous orbit? outside which, the torques on their orbital motions 
are positive. One possibility is of course gas drag.  Another 
possibility is that somehow as the star  rotation speed 
decreases its synchronous orbit moves outwards leaving these planets
in a new position inside the synchronous orbit. 
Last but not least important is to consider 
the possibility that these planets formed {\it in situ}, that is, very close
to where they are found now and therefore their chemical composition has to be
dominated (by a large quantity) of very refractory elements or compounds.
Venus has the largest albedo in the Solar System due to its atmosphere which is
mainly composed by carbon dioxide. Hence the chemical composition of these
planets can have a large albedo and a large abundance of heavy elements 
that did not moved far from the star due the high inner temperatures 
in the disk,
and hence may have formed these planets that have large gravitational fields 
such that they can retain their heavy materials. 
Nevertheless their unavoidable fate
is to fall onto the star because even if gas drag is still present, its
effect works in the same
direction as the tidal force on the planet caused by the lagging bulge it 
raises on the star.

I would like to thank the Instituto de Astronomía UNAM where this
project started and The Facultad de Ciencias UNAM for their great
support and encouragement. Also I am very grateful to Dr. Edmundo Moreno
for his insightful comments and questions and to Dr. Arcadio Poveda
and Dr. Ramón Peralta y Fabi for our very enlighting discussions.
  
\linespread{1.}
\begin{table*}[p]\centering
  \setlength{\tabcolsep}{0.5em}
\begin{center}
    \caption{Hosts star data}
    \label{tab:1}
    \begin{tabular}{lllcccccccr}\hline\hline
Star   &  Identifier  & Spectral    & Mass & Luminosity & $T_{eff}$ &
[Fe/H] & P$_{rot}$ & $R_*$& Age& Number of \\
No. & & Type &  $M_\odot$   & $L_\odot$ & K & & days& R$_\odot$& Gyr&planets \\
\hline

 1&OGLETR56       &G  - -&  1.04& -   &  -   &  -   &   -  & 1.12 &  - &  1 \\
 2&OGLETR113      &K  - -&  0.77& -   &  -   &  0.14&   -  & 0.765&  - &  1 \\
 3&OGLETR132      &F  - -&  1.35& -   &  -   &  0.43&   -  & 1.43 & 1.4&  1 \\
 4&HD73256$^5$    &K  0 V&  1.05& 0.69&  5570&  0.29& 13.9 & 1.03 &0.83&  1 \\
 5&GJ436          &M 2.5V&  0.41&0.025&  -   &  0.25& -    & 0.43 &  - &  1 \\
 6&HD75732$^5$    &G 8 V & 0.95 & 0.61&  5250&  0.16&  38.5& 0.96 & 5.0&  4 \\
 7&HD63454$^5$    &K 4 V & 0.8  & 0.26&  4841&  0.11& -    & 0.84 &  - &  1 \\
 8&HD83443$^5$    &K 0 V & 0.9  & 0.88&  5454&  0.35&  35.3& 0.92 & 3.2&  1 \\
 9&HD46375$^1$    &K 1 V &  1.0 &  1.0&  5770&  0.34& -    & 1.0  &  - &  1 \\
 10&TrES-1$^6$    &K 0 V &  0.87&  0.5&  5250& 0.001& -    & 0.85 &  - &  1 \\
 11&HD179949$^2$  &F 8 V &  1.24& 1.99&  6155&  0.02&   9. & 1.24 &  - &  1 \\
 12&HD187123      &G 3 V &  1.06& 1.35&  5830&  0.16&  25.4& 1.18 &  - &  1 \\
 13&OGLE-TR-10    &G - - &  1.22&  -  &  -   &  0.12& -    & -    &  - &  1 \\
 14&HD120136$^2$  &F 8 V &  1.3 & 2.31&  6498&  0.28&  3.3 & 1.2  & 2.0&  1 \\
 15&HD330075      &K 1 - &  0.7 & 0.47&  5017&  0.08&  48  & -    &6.2 & 1 \\
 16&HD88133$^+$   &G 5 IV&  1.2&  3.06&  5494&  0.34&  48  & 1.93 &  - & 1 \\
 17&HD2638        &G 5 -&  0.93&  0.47&  5192&  0.16&  37  &     -&  - & 1 \\
 18&BD103166$^6$  &K 0 V&   1.1&  0.62&  5400&  0.50&   -  &   0.9&  - & 1 \\
 19&HD75289$^5$   &G 0 V&  1.15&  1.99&  6000&  0.29& 15.95&  1.08& 5.6& 1 \\
 20&HD209458$^5$  &G 0 V&  1.03&  1.61&  6025&  0.04&  14.4&  1.02&  5.& 1 \\
 21&HD76700$^4$   &G 8 V&  1   &  1   &  5423&  0.14&   -  &  1   & -  & 1 \\
 22&OGLETR111$^+$ &G - -&  0.82&  0.43&  5070&  0.12&   -  &  0.85&  - & 1 \\
 23&HD217014$^1$  &G 5 V&  1.06&   1.2&  5946&  0.20&  28. &  0.03&  - & 1 \\
 24&HD9826        &F 8 V&   1.3&   3.4&  6210&   0.1&  10.2&  1.4 &  - & 3 \\
 25&HD49674$^1$   &G 5 V&   1  &   1.0&  5770&  0.25&  27.2&    1 &  - & 1 \\
 26&HD68988$^1$   &G 2 V&   1.2&  1.79&  6338&  0.24&  26.7&  1.1 & 6. & 1 \\
 27&HD168746      &G 5 -&  0.88&   1.1&  5610& -0.06& -    &  -   & -  & 1 \\
 28&HD217107$^4$  &G 7 V&  0.98&  0.94&  5700&  0.32&  39  &0.98  &7.76& 1 \\
 29&HD162020$^5$  &K 2 V&  0.75&  0.25&  4830&  0.01&     -&0.79  & -  & 1 \\
 30&HD160691$^5$  &G 5 V&  1.1 &  1.77&  5813&  0.32&  31  &1.05  & 2. & 3 \\
 31&HD130322$^5$  &K 0 V&  0.79&   0.5&  5330& -0.02&  8.7 &0.83  &0.35& 1 \\
 32&HD108147$^5$  &G 0 V&  1.27&  1.93&  6265&  0.20&  8.7 &1.15  &2.17& 1\\
 33&HD38529$^+$   &G 4IV&  1.39&  5.96&  5370&  0.35&  34.5&  2.82&  - & 2\\
 34&HD13445$^5$   &K 0 V&  0.8 &  0.4 &  5350& -0.24&  31  &  0.84&  - & 1 \\
 35&HD99492$^6$   &K 2 V&  0.88&  0.33&  4954&  0.36& -    &  0.79&  - & 1 \\
 36&HD27894$^5$   &K 2 V&  0.75&  0.36&  4875&  0.3 & -    &  0.79&  - & 1 \\
 37&HD195019$^4$  &G 3 V&  1.02&  1.06&  5600&  0.0 &  24.3&  1.01&3.16& 1 \\
 38&HD6434$^5$    &G 3 V&  0.79&  1.12&  5835& -0.52&  18.6&  0.83&3.8 & 1\\
 39&HD192263$^5$  &K 2 V&  0.75&  0.34&  4840& -0.14&  9.5 &0.79  &  -  &1 \\
 40&Gl876$^5$     &M 4 V&  0.3&  0.014&  3200&  0.0 & -    &  0.38&  -  & 2$^*$ \\
 41&HD102117$^5$  &G 6 V&  1.03&  1.57&  5672&  0.3 &  34  &  1.02&  -  & 1 \\
  \hline\hline
 
      \end{tabular}
  \end{center}
\end{table*}
\linespread{1.}

\linespread{1.}
\begin{table*}[p]\centering
  \setlength{\tabcolsep}{0.7em}
\begin{center}
    \caption{Hosts star data continues}
    \label{tab:1a}
    \begin{tabular}{lllcccccccr}\hline\hline
Star   &  Identifier  & Spectral    & Mass & Luminosity & $T_{eff}$ &
[Fe/H] & P$_{rot}$ & $R_*$& Age& Number of \\
No. & & Type &  $M_\odot$   & $L_\odot$ & K & & days& $R_\odot$& Gyr&planets \\
\hline
 42&HD11964       &G 5 -& 1.125&  -   &  -   & -    & -  &  -   &  -   & 2 \\
 43&HD143761$^6$  &G 0 V&  0.95&  1.94&  5860& -0.19&  20&  1.35&  10. & 1 \\
 44&HD74156       &G 0 -&  1.27&  3.06&  6105&  0.15& -  &  -   &  -   & 2 \\
 45&HD117618$^6$  &G 2 V&  1.05&  1.8 &  5860&  0.04& -  &  1.3 &  -   & 1 \\
 46&HD37605$^6$   &K 0 V&  0.8 &  0.74&  5475&  0.39& -  &  0.96&  -   & 1 \\
 47&HD168443$^+$  &G 5IV&  1.01&  1.93&  5555&  0.03&26.8&  1.5 &  2.7 & 2 \\
 48&HD3651$^1$    &K 0 V&  0.79&  0.47&  5251&  0.05&44.5&  0.83&  -   & 1 \\
 49&HD219449      &K0III& -    &  -   &  -   &  0.05& -  &  -   &  -   & 1 \\
 50&HD101930      &K 1 V&  0.74&  0.49&  5079&  0.17&  46&  0.93&  -   & 1 \\
 51&HD121504      &G 2 V&  1.18&  1.55&  6075&  0.16& 8.6&  1.27&  1.2 & 1 \\
 52&HD178911B     &G 5 V&  0.87&  1.3 &  5650&  0.28& -  &  1.14&  -   & 1 \\
 53&HD16141$^+$   &G 5IV&  1.01&  2.46&  5770&  0.22& -  &  1.57&  -   & 1 \\
 54&HD114762$^6$  &F 9 V&  0.82&  1.75&  6110& -0.7 & -  &  1.18&  -   & 1 \\
 55&HD80606       &G 5 V&  1.1 &  0.75&  5645&  0.43& -  &  0.86&  -   & 1 \\
 56&HD216770      &K 0 V&  0.9 &  0.79&  5229&  0.23&35.6&  1.1 &  3.1 & 1 \\
 57&HD93083       &K 3 V&  0.7 &  0.41&  4995&  0.15&  48&  1.03&  -   & 1 \\
 58&HD117176$^6$  &G 5 V&  1.1 &  3.1 &  5770& -0.03&34.8&  1.76&  8.2 & 1 \\
 59&HD52265       &G 0 V&  1.18&  1.98&  6060&  0.21&14.6&  1.35&  4.  & 1 \\
 60&HD208487$^6$  &G 2 V&  1.3 &  1.82&  5860& -0.06& -  &  1.31&  -   & 1 \\
 61&HD1237$^5$    &G 6 V&  0.9 &  0.66&  5540&  0.1 &12.6&  0.92&  0.8 & 1 \\
 62&HD34445       &G 0 -&  1.1 &  -   &  -   &  0.14& -  &  -   &  -   & 1 \\
 63&HD37124$^1$   &G 4 V&  0.91&  0.74&  5556& -0.42&25  & 0.93 &  3.9 & 3 \\
 64&HD73526$^6$   &G 6 V&  1.02&  2.2 &  5700&  0.28& -  &  1.52&  -   & 1 \\
 65&HD104985      &G 9 3&  1.5 &  -   &  5410& -0.35& -  &  8.73&  -   & 1 \\
 66&HD82943$^+$   &G 0 -&  1.15&  1.63&  6028&  0.29&  18&  1.17&2.9   & 2 \\
 67&HD169830      &F 8 V&  1.4 &  4.59&  6299&  0.21& 8.3&  1.95&  2.8 & 2 \\
 68&HD8574        &F 8 V&  1.17&  2.25&  6080&  0.05& -  &  1.35&  -   & 1 \\
 69&HD202206      &G 6 V&  1.15&  1.07&  5765&  0.37& 9.5&  1.13&   5.6& 2 \\
 70&HD89744$^6$   &F 7 V&  1.4 &  6   &  6166&  0.18& 9  &  2.14& 2.04 & 1 \\
 71&HD134987$^1$  &G 5 V&  1.05&  1.17&  5917&  0.23&30.34&  1.02&  -  & 1 \\
 72&HD40979$^6$   &F 8 V&  1.1 &  1.97&  6095& 0.194&  7  &  1.26& 1.5 & 1 \\
 73&HD12661$^6$   &G 6 V&  1.07&  1.19&  5754&  0.29 & 36& 1.096&1.47  & 2 \\
 74&HD150706      &G 0 V&  0.98&  0.98&  5886& -0.13& -  &  0.93&  -   & 1 \\
 75&HD59686       &K2III&  1.1 &  -   &  4871& 0.28 & -  &  -   &  -   & 1 \\
 76&HD17051       &G 0 V&  1.03&  1.52&  6125&  0.11&  8 &  1.097&1.556 & 1 \\
 77&HD142$^+$     &G1 IV&  1.15&  3.24&  6302&  0.14& -  &  1.7  &  -   & 1 \\
 78&HD92788$^+$   &G 5 -&  1.1 &  1.18&  5821&  0.34&21.3&  1.07 &  2.1 & 1 \\
 79&HD28185       &G  5V&  0.99&  1.02&  5705&  0.24&  30&  1.05 & 2.9  & 1 \\
 80&HD196885      &F 8IV&  1.27&  -   &  -   &  0.2 & -  &  -    &  -   & 1 \\
 81&HD142415      &G 1 V&  1.03&  1.14&  6045&  0.21&  9.6&  1.06& 1.1  & 1 \\
 82&HD177830      &K 0IV&  1.15&  -   &  -   &  0.0 &70.94&  -   &  -   & 1 \\
 \hline\hline

       \end{tabular}
  \end{center}
\end{table*}
\linespread{1.}

\linespread{1.}
\begin{table*}[p]\centering
  \setlength{\tabcolsep}{0.7em}
\begin{center}
    \caption{Hosts star data continues}
    \label{tab:1b}
    \begin{tabular}{lllcccccccr}\hline\hline
Star   &  Identifier  & Spectral    & Mass & Luminosity & $T_{eff}$ &
[Fe/H] & P$_{rot}$ & $R_*$& Age& Number of \\
No. & & Type &  $M_\odot$   & $L_\odot$ & K & & days& $R_\odot$& Gyr&planets \\
\hline
 83&HD154857$^6$  &G 5 V&  1.17&  4.95&  5628& -0.23& -     &  2.34&5. & 1 \\
 84&HD108874$^6$  &G 5 V&  1   &  1.34&  5770&  0.23&  38   &  1.16& - &2$^*$ \\
 85&HD4203$^1$    &G 5 V&  1.06&  1.2 &  5946&  0.22&  43.1 &  1.03& - & 1 \\
 86&HD128311$^6$  &K 0 V&  0.8 &  0.28&  5250&  0.03&  14   &  0.64& 1.99 & 2 \\
 87&HD27442       &K 2 IV&  1.2&  -   &  -   &  0.22& -     &  -   &  10. & 1 \\
 88&HD210277      &G 7 V&  0.92&  0.93&  5570&  0.24&  40.8 &  0.91& 8.5  & 1 \\
 89&HD19994$^5$   &F 8 V&  1.34&  3.81&  6121&  0.19&  12.2 &  1.18& 2.4  & 1 \\
 90&HD188015$^+$  &G 5 IV&  1.08&  1.44&  5745&  0.29&  36  &  1.21&  9.82& 1\\
 91&HD20367$^+$   &G 0 -&  1.17 &  1.88&  6100&  0.14& -    &  1.23&  -   & 1 \\
 92&HD114783$^6$  &K 2 V&  0.92 &  0.4 &  5250&  0.33&  45.2&  0.76&  -   & 1 \\
 93&HD147513$^5$  &G 5 V&  1.11 &  0.98&  5883&  0.06&  4.7 &  1.06&  0.3 & 1 \\
 94&HD137759      &K 2 3&  1.05 &  -   &  4900&  0.03& -    &  11.01&  -  & 1 \\
 95&HD222582$^6$  &G 5 V&  1    &  1.14&  5770& -0.01& -    &  1.07 &  -  & 1 \\
 96&HD65216       &G 5 V&   0.92&  0.71&  5666& -0.12& -    &   0.87&  -  & 1\\
 97&HD183263$^+$  &G 2IV&  1.17 &  2.04&  5936&  0.3 &  32  &   1.35&9.91 & 1 \\
 98&HD141937      &G 2 V&  1.1  &  1.17&  5925&  0.11& 13.25&  1.1  & 1.6 & 1 \\
 99&HD41004A$^5$  &K 2 V&  0.7  &  0.65&  5010&  0.16&  27  & 0.75  &1.6  & 1$^*$ \\
 100&HD47536      &K 0 3&  2.05 &  -   &  4554& -0.54& -    &  -    &  -  & 1\\
 101&HD23079$^6$  &G 0 V&  1.1  &  1.5 &  6200& -0.11& -    &  1.06 &  -  & 1 \\
 102&HD186427$^4$ &G 5 V&  1    &  1   &  5760&  0.05&  29.1&  1    &  -  & 1 \\
 103&HD4208$^1$   &G 5 V&  0.93 &  0.79&  5605& -0.24&  25.3&  0.94 &  -  & 1 \\
 104&HD114386$^5$ &K 3 V&  0.68 &  0.29&  4804&  0.09& -    &  0.73 &  -  & 1 \\
 105&HD45350$^6$  &G 5 V&  1.02 &  1.38&  5616&  0.29&  39  &  1.24 &  9.9& 1 \\
 106&HD222404     &K 1IV&  1.59 &  -   &  4888&  0.0 & -    &  4.66 &  -  & 1 \\
 107&HD213240$^+$ &G 4IV&  1.22 &  3.6 &  5975&  0.16&  15  &  1.77 &  2.7& 1 \\
 108&HD10647      &F 8 V&  1.07 &  1.51&  6143& -0.03&  7.2 &  1.11 &1.75 & 1 \\
 109&HD10697$^+$  &G 5IV&  1.1  &  3.42&  5770&  0.15&  32.6&  1.85 &  -  & 1\\
 110&HD95128$^6$  &G 0 V&  1.03 &  1.11&  5780&  0.01&  24  &  1.05 &  7. & 2 \\
 111&HD190228$^+$ &G 5IV&  0.83 &  3.6 &  5360& -0.24& -    &  2.2  &  -  & 1\\
 112&HD114729$^6$ &G 0 V&  0.93 &  2.16&  5915& -0.22& 21.58&  1.4  &  6. & 1 \\
 113&HD111232     &G 5 V&  0.78 &  0.69&  5494& -0.36&  30.6&  0.84 &  5.2& 1 \\
 114&HD2039$^6$   &G 2.5V&  0.98&  1.73&  5675&  0.1 & -    &  1.36 &  -  & 1\\
 115&HD136118$^6$ &F 9 V&  1.24 &  2.92&  6003&-0.065&  12.2&  1.58 &  3. & 1 \\
 116&HD50554      &F 8 V&  1.11 &  1.45&  6050&  0.02&  16.1&  1.08 & 4.5 & 1\\
 117&HD196050     &G 3 V&  1.1  &  1.83&  5918&  0.22&  16  &  1.39 & 1.6 & 1\\
 118&HD216437$^5$ &G 4 V&  1.06 &  2.25&  5887&  0.25&  26.7&  1.03 &  5.8& 1 \\
 119&HD216435$^6$ &G 0 V&  1.25 &  4.25&  5830&  0.15& -    &  2.02 &  5. & 1\\
 120&HD106252     &G 0 V&  1.02 &  1.27&  5890& -0.01&  22.8&  1.07 &  5. & 1 \\
 121&HD23596$^+$  &F 8 -&  1.3  &  2.86&  6125&  0.32& -    &  1.5  &  -  & 1 \\
 122&HD145675$^5$ &K 0 V&  0.9  &  0.71&  5255&  0.51&  41  &  0.92 & 3.9 & 1 \\
 123&HD142022     &K 0 V&  0.99 &  1.01&  5500&  0.19&  38  &  1.25 &11.5 & 1 \\
  \hline\hline

       \end{tabular}
  \end{center}
\end{table*}

\begin{table*}[p]\centering
  \setlength{\tabcolsep}{0.7em}
\begin{center}
    \caption{Hosts star data continues}
    \label{tab:1d}
    \begin{tabular}{lllcccccccr}\hline\hline
Star   &  Identifier  & Spectral    & Mass & Luminosity & $T_{eff}$ &
[Fe/H] & P$_{rot}$ & $R_*$& Age& Number of \\
No. & & Type &  $M_\odot$   & $L_\odot$ & K & & days& $R_\odot$& Gyr&planets \\
\hline
 124&HD39091$^1$  &G 1 V&  1.1 &  1.36&  6060&   0.09&     -&  1.05&  - &1 \\
 125&HD72659$^6$  &G 0 V&  0.95&  2.44&  6030& -0.14 &  20.3&  1.43&  - &1 \\
 126&HD70642$^6$  &G 5 V&  1.0 &  1.34&  5670&   0.16&     -&  1.2 &  4.&1\\
 127&HD33636      &G 0 V&  1.12&  1.07&  5990&  -0.05&  14.3&  0.99&2.8 &1 \\
 128&HD22049$^6$  &K 2 V&  0.8&  0.47&  5180&    -0.1&    12&  0.85&0.7 &1\\
 129&HD117207$^6$ &G 8 V&  1.04&  1.68&  5723&  0.27&     36&  1.32&9.84&1 \\
 130&HD30177$^6$  &G 8 V&  0.95&  1.18&  5320&   0.2&     - &  1.28&  - &1 \\
 131&HD50499$^1$  &G 1 V&  1.27&  2.15&  6526&   0.3&    21 &  1.15&  - &1 \\
 132&HD190360$^+$ &G 6 IV&  0.96&  1.27&  5590& 0.24&     40&   1.2&  6.7&2 \\
 133&HD89307$^1$  &G 0 V&  1.27&  2.15&  6526& -0.16&     -&   1.15&  -  &1 \\

  \hline\hline
       \end{tabular}
  \end{center}
\end{table*}

\linespread{1.}
\begin{table*}[p]\centering
  \setlength{\tabcolsep}{0.6em}
\begin{center}
    \caption{Planet data}
    \label{tab:2}
    \begin{tabular}{lllccccccll}\hline\hline

Planet& Identifier & Mass & R$_p$ & a & e &P$_{orb}$ & $r_{av}$ &I       &r$_{ap}$/r$_{s}$&i \\
No.   &            & $M_J$& $R_J$ &AU &   &days      & UA       &$10^6$erg/cm$^{2}$s&     &deg \\
\hline
 
 1&OGLETR56b &1.45 & 1.23& 0.0225&  0.  & 1.212& 0.0225&  -  &  -  &81.0 \\
 2&OGLETR113b&1.35 & 1.08& 0.0228&  0.  & 1.43 & 0.0228&  -  &  -  & -   \\
 3&OGLETR132b&1.01 & 1.15& 0.0306&  0.  & 1.69 & 0.0306&  -  & -   & -  \\
 4&HD73256b  &1.87 &  -  & 0.037 &0.029 & 2.55 & 0.0370&5.3  &0.331& -  \\
 5&GJ436b    &0.067&  -  &  0.028& 0.12 & 2.644& 0.0282& 42.8&  -  & - \\
 6&HD75732e  &0.045&  -  &  0.038& 0.174& 2.81 & 0.0386&557.8&0.203& - \\
 7&HD75732b  & 0.84&  -  &  0.115&  0.02&14.65 & 0.115 &62.7 &0.535& - \\
 8&HD75732c  & 0.21&  -  &  0.241& 0.339&44.276& 0.2548&12.78&1.471& - \\
 9&HD75732d  & 4.05&  -  &  5.9  & 0.16 &5360. & 5.9752&0.023&31.21& - \\
 10&HD63454b & 0.38&  -  &  0.036& 0.   & 2.818& 0.036 &272.9&  - & - \\
 11&HD83443b & 0.38&  -  &  0.039&0.013 & 2.985& 0.0390&787.2&0.194& -  \\
 12&HD46375b & 0.25&  -  &  0.041& 0.04 & 3.024& 0.0410&808.2&  -  & - \\
 13&TrES-1b  & 0.75& 1.08& 0.0393&0.135 &  3.03& 0.0396&428.3&  -  & 88.2  \\
 14&HD179949b& 0.98&  -  &  0.04 &  0.  & 3.093& 0.04  &1692.8&0.439& - \\
 15&HD187123b& 0.52&  -  &  0.042& 0.03 & 3.097& 0.0420&1040.5&0.251& - \\
 16&OGLE-TR-10b&0.57&1.24& 0.0416&  0.  & 3.101& 0.0416&  -   &  -  & 98.2 \\
 17&HD120136b& 4.13&  -  &  0.05 &  0.01& 3.312& 0.050 &1260.1&1.067& - \\
 18&HD330075b& 0.62&  -  &  0.039&  0.  & 3.388& 0.039 &420.5 &0.170& - \\
 19&HD88133b&  0.29&  -  &  0.046& 0.11 &  3.41& 0.046 &1944.6&0.186& - \\
 20&HD2638b &  0.48&  -  &  0.044&  0.  & 3.444& 0.044 &330.3 &0.207& - \\
 21&BD103166b& 0.48&  -  &  0.046& 0.05 & 3.487& 0.046 &1515.8&  -  & 84.3  \\
 22&HD75289b & 0.42&  -  &  0.046&  0.  & 3.509& 0.046 &1279.7&0.354& - \\
 23&HD209458b& 0.67& 1.43&  0.05 & 0.11 & 3.525& 0.050 &865.8 &0.474& 86.1 \\
 24&HD76700b & 0.19&  -  &  0.05 & 0.13 & 3.971& 0.0504&417.5 &  -  & - \\
 25&OGLETR111b&0.53&  1. &  0.047&  0.  & 4.0161& 0.047&265.2 &  -  & 86.5 \\
 26&HD217014b& 0.45&  -  &  0.05 &  0.01& 4.231 & 0.050&655.8 &0.274& - \\
 27&HD9826b&  0.69 &  -  &  0.059& 0.012& 4.617 & 0.059&1330.0&0.594& -  \\
 28&HD9826c&  1.89 &  -  &  0.829& 0.28 & 241.5 & 0.861&6.2   &10.565& - \\
 29&HD9826d&  3.75 &  -  &  2.53 &  0.27&  1284 & 2.622&0.673 &31.992& - \\
 30&HD49674b&  0.11&  -  &  0.06 &  0.16&  4.95 & 0.061&368.5 & 0.393& -  \\
 31&HD68988b&  1.92&  -  &  0.07 &  0.15&  6.276& 0.071&486.7 & 0.433& - \\
 32&HD168746b& 0.23&  -  &  0.065&  0.081& 6.403& 0.0652&351.9&  -   & - \\
 33&HD217107b& 1.37&  -  &  0.074&  0.13 & 7.126& 0.0746&245.9& 0.366& - \\
 34&HD162020b& 14.4&  -  &  0.074&  0.277& 8.428& 0.0768&57.6 &  -   & - \\
 35&HD160691d&0.044&  -  &  0.09 &  0.   &  9.55& 0.09  &297.3& 0.451& - \\
 36&HD160691b& 1.67&  -  &  1.5  &  0.2  & 654.5& 1.53  &1.0  & 9.031& - \\
 37&HD160691c&  3.1&  -  &  4.17 &  0.57 & 2986.& 4.85  & 0.1 &32.847& - \\
 38&HD130322b& 1.02&  -  &  0.088&  0.044& 10.72& 0.09  &87.7 & 1.201& - \\
 39&HD108147b&  0.4&  -  &  0.104&  0.498& 10.90& 0.12  &192.2& 1.738& - \\
 40&HD38529b & 0.78&  -  &  0.129&  0.28 & 14.31& 0.134 &451.6& 0.713& - \\
 41&HD38529c & 12.8&  -  &  3.71 &  0.33 &2207.4& 3.912 &0.53 &21.322& - \\
 \hline\hline
       \end{tabular}
  \end{center}
\end{table*}
\linespread{1.}

\linespread{1.}
\begin{table*}[p]\centering
  \setlength{\tabcolsep}{0.6em}
\begin{center}
    \caption{Planet data}
    \label{tab:2}
    \begin{tabular}{lllccccccll}\hline\hline

Planet& Identifier & Mass & R$_p$ & a & e &P$_{orb}$ & r$_{av}$ &I                  &r$_a$/r$_s$&i \\
No.   &            & M$_J$& R$_J$ & AU&   &days      & UA       &10$^6$erg/cm$^{2}s$&           &deg\\
\hline

 42&HD13445b&  4. & -   &  0.11&  0.046 & 15.78 & 0.1101&44.9  & 0.64& -  \\
 43&HD99492b& 0.12& -   & 0.124&  0.05  & 17.00 & 0.1241&29.9  & -   & -  \\
 44&HD27894b& 0.62& -   & 0.122&  0.049 & 17.99 & 0.1221&32.8  & -   & - \\
 45&HD195019b&3.57& -   &  0.14&  0.02  & 18.20 & 0.1400&62.9  & 0.86& -  \\
 46&HD6434b&  0.39& -   &  0.14&  0.17  & 21.99 & 0.1420&75.5  &1.29 & -  \\
 47&HD192263b&0.73& -   &  0.15&  0.    & 24.13 & 0.15  &20.6  &1.88&  - \\
 48&Gl876b&   0.56& -   &  0.13&  0.27  & 30.12 & 0.1347&1.05  & -  & - \\
 49&Gl876c&  1.98 & -   &  0.21&  0.12  & 61.02 & 0.2115&0.426 & -  & 84 \\
 50&HD102117b&0.14& -   &  0.15&  0.    & 20.67 & 0.15  &94.9  &0.72 & - \\
 51&HD11964b& 0.11& -   & 0.229&  0.15  & 37.82 & 0.2316& -    & -   & -  \\
 52&HD11964c&  0.7& -   & 3.167&  0.3   &  1940.& 3.3095& -    & -   & -  \\
 53&HD143761b&1.04& -   &  0.22&  0.04  & 39.945& 0.2202&54.4  &1.61 & - \\
 54&HD74156b& 1.86& -   & 0.294&  0.636 & 51.643& 0.3535&33.3  & -   & - \\
 55&HD74156c& 6.17& -   &  3.4 &  0.583 & 2025. & 3.9778&0.3   & -   & -  \\
 56&HD117618b&0.22& -   &  0.28&  0.29  &  52.16& 0.2918&28.7  & -   & - \\
 57&HD37605b &2.85& -   &  0.26&  0.736 &  54.2 & 0.3304&9.3   & -   & - \\
 58&HD168443b&7.73& -   & 0.295&  0.53  &  58.1 & 0.3364&23.2  &2.57 & - \\
 59&HD168443c&17.2& -   &  2.87&  0.2   & 1739.5& 2.9274&0.3   &19.59& - \\
 60&HD3651b&   0.2& -   & 0.284&  0.63  &  62.23& 0.3403&5.5   & 2.04& - \\
 61&HD219449b& 2.9& -   &  0.3 &   -    &  182. & -     & -    & -  & - \\
 62&HD101930b& 0.3& -   & 0.302&  0.11  &  70.46& 0.3038& 7.2  &1.47& - \\
 63&HD121504b&1.22& -   & 0.33 &  0.03  &  63.33& 0.3301&19.3  &3.91 & - \\
 64&HD178911Bb&6.292& - & 0.32 &  0.1243& 71.487& 0.3225&17.0  & -   & - \\
 65&HD16141b& 0.23& -   & 0.35 &  0.28  &  75.82& 0.3637& 25.3 & -   & - \\
 66&HD114762b&11.03& -  & 0.35 &  0.34  & 83.895& 0.3702& 17.4 & -   & -  \\
 67&HD80606b&  3.9&  -  & 0.469&  0.927 &111.81 & 0.6705&2.3   & -   & - \\
 68&HD216770b&0.65& -   & 0.46 &  0.37  &118.45 & 0.4914&4.4   &3.08 & - \\
 69&HD93083b &0.37& -   & 0.477&  0.14  &143.58 & 0.4817&2.4   &2.37 & -  \\
 70&HD117176b&7.44& -   &  0.48&  0.4   &116.689& 0.5184&15.7  &3.12 & - \\
 71&HD52265b&  1. & -   &  0.49&  0.35  &119.1  & 0.5200&9.9   &5.36 & - \\
 72&HD208487b&0.45& -   &  0.49&  0.32  &  130. & 0.5151&9.4   & -   & - \\
 73&HD1237b&  3.37& -   &  0.49&  0.511 & 133.71& 0.5540&2.9   &7.24 & - \\
 74&HD34445b& 0.58& -   &  0.51&  0.4   & 126.  & 0.5508& -    & -   & - \\
 75&HD37124b& 0.61& -   &  0.53&  0.055 & 154.46& 0.5308&2.2   &3.58 & - \\
 76&HD37124c& 0.6 & -   &  1.64&  0.14  &  843.6& 1.6572&0.2   &11.96& - \\
 77&HD37124d&0.66 & -   &  3.19&  0.2   &  2295.& 3.2538&0.06  &24.48& -  \\
 78&HD73526b&2.98 & -   &  0.65&  0.44  &184.108& 0.7129&5.89  & -   & -  \\
 79&HD104985b&6.3 & -   &  0.78&  0.03  &198.2  & 0.7803& -    & -   & - \\
 80&HD82943c &1.85& -   &  0.75&  0.38  &  219. & 0.80415&3.43 &7.35 & -  \\
 81&HD82943b& 1.84& -   &  1.18&  0.18  &  435. &  1.1991&1.54 &9.89 & - \\
 82&HD169830b&2.88& -   &  0.81&  0.31  & 225.62&  0.8489&8.67 &11.82& - \\

  \hline\hline
       \end{tabular}
  \end{center}
\end{table*}
\linespread{1.}

\linespread{1.}
\begin{table*}[p]\centering
  \setlength{\tabcolsep}{0.6em}
\begin{center}
    \caption{Planet data}
    \label{tab:2}
    \begin{tabular}{lllccccccll}\hline\hline
Planet& Identifier & Mass & R$_p$ & a & e &P$_{orb}$ & r$_{av}$ &I            &r$_a$/r$_s$    &i \\
No.   &            & M$_J$& R$_J$ &AU &   &days      & UA       &10$^6$erg/cm$^{2}$s&         & deg \\
\hline

 83&HD169830c&  4.04& -&  3.6&  0.33 &  2102.&  3.7960&0.4334&53.4& - \\
 84&HD8574b&    2.11& -& 0.77& 0.288 & 227.55& 0.8019 &4.76  &   -& - \\
 85&HD202206b&  17.5& -& 0.83& 0.433 &  256.2& 0.9078 &1.77  &12.9& - \\
 86&HD202206c&  2.41& -& 2.44& 0.284 & 1296.8& 2.5384 &0.23  &34  & - \\
 87&HD89744b&  7.99 & -& 0.89& 0.67  &  256.6& 1.0898 &6.84  &15.7& - \\
 88&HD134987b& 1.58 & -& 0.81& 0.24  &  259. & 0.8333 &2.29  &5.2 & - \\
 89&HD40979b&  3.32 & -& 0.81& 0.23  &  267.2& 0.8324 &3.88  &13.4& - \\
 90&HD12661b&  2.3  & -& 0.82& 0.33  &  263 & 0.8646 &3.89   &5.0 & - \\
 91&HD12661c&  1.5  & -& 2.6 & 0.2   &  1530& 2.652  &0.41   &14.3& - \\
 92&HD150706b& 1    & -& 0.82& 0.38  & 264.9& 0.8792 &1.72   &-   & -  \\
 93&HD59686b&  5.25 & -& 0.91& -     &  303 & 0.91   & -     & -  & -  \\
 94&HD17051b&  1.94 & -& 0.91& 0.24  &311.288&0.9362 &2.36  &14.3& -  \\
 95&HD142b&     1.07& -& 0.97& 0.37  &331.872&1.03639&3.17  & -  & - \\
 96&HD92788b&  3.58 & -& 0.96& 0.35  & 325   &1.0188 &1.55  & 8.3& - \\
 97&HD28185b&  5.7  & -& 1.03&  0.07 & 383   &1.0325 &1.30  & 5.8&  -\\
 98&HD196885&  1.84 & -& 1.122&  0.3 & 386   &1.17249& -    & -  & - \\
 99&HD142415b& 1.62 & -& 1.05 & 0.5  & 386.3 &1.18125&1.11  &17.6& - \\
 100&HD177830b&1.52 & -& 1.14 & 0.1  & 408.377&1.1457& -    &3.6 & -  \\
 101&HD154857b&1.8  & -& 1.11 & 0.51 & 398    &1.2543&4.28  & -  & -  \\
 102&HD108874b&1.36 & -& 1.051& 0.07 & 395.4  &1.0535&1.65  &5.1 & -  \\
 103&HD108874c&1.02 & -& 2.68 & 0.25 &1605.8  &2.7637&0.24  &15.2&  - \\
 104&HD4203b  &3.35 & -& 1.09 & 0.51 & 404.224&1.2317&1.08  & 6.7& -  \\
 105&HD128311b& 2.57& -& 1.02 & 0.31 & 422    &1.0690&0.27  &12.4& - \\
 106&HD128311c& 2.18& -& 1.1  & 0.25 & 458.6  &1.1344&0.24  &12.8& -  \\
 107&HD27442b&  1.28& -&  1.16&0.058 & 415.2  &1.1619& -    & -  & - \\
 108&HD210277b&  1.3& -&  1.12& 0.46 & 434.289&1.2385&0.82  &7.2  &- \\
 109&HD19994b&  1.68& -&  1.42&  0.3 & 535.7  &1.4839&2.35  &16.1 & - \\
 110&HD188015b& 1.26& -&  1.19&  0.15& 456.5  &1.2034&1.35  & 6.2 & - \\
 111&HD20367b&  1.17& -&  1.25&  0.32& 469.5  &1.314 &1.49  & -   & -  \\
 112&HD114783b& 1   & -&  1.2 &  0.1 & 501    &1.206 &0.37  &5.46 & - \\
 113&HD147513b& 1.21& -&  1.32&  0.26& 528.4  &1.3646&0.71  &29.2& - \\
 114&HD137759b& 8.47& -&  1.28&  0.72& 510.833&1.6117& -    & -  & - \\
 115&HD222582b& 5.11& -&  1.35&  0.76& 572.   &1.7399&0.51  & -  & - \\
 116&HD65216b & 1.21& -&  1.37&  0.41& 613.1  &1.4851&0.44  & -  & - \\
 117&HD183263b& 3.69& -&  1.52&  0.38& 634.23 &1.6297&1.04  &10.0&  -\\
 118&HD141937b&  9.7& -&  1.52&  0.41& 653.22 &1.6477&0.59  &19.0& - \\
 119&HD41004Ab&2.436& -&  1.64&  0.5 &  924.  &1.845 &0.26  &15.7& - \\
 120&HD41004Bb&18.37& -&  0.02&  0.081&1.33   &0.0201&2196.8&0.1 & -  \\
 121&HD47536b &7.315& -&  1.93&  0.2  & 712.13&1.9686& -     & - & - \\
 122&HD23079b &  2.5& -&  1.5 &  0.04 & 738.46&1.5012&0.90   & - & - \\
 123&HD186427b& 1.69& -&  1.67&  0.67 &798.94 &2.0448&0.32   &15.1& - \\

  \hline\hline
       \end{tabular}
  \end{center}
\end{table*}
\linespread{1.}

\linespread{1.}
\begin{table*}[p]\centering
  \setlength{\tabcolsep}{0.5em}
\begin{center}
    \caption{Planet data}
    \label{tab:2}
    \begin{tabular}{lllccccccll}\hline\hline
Planet& Identifier & Mass & R$_p$ & a & e &P$_{orb}$ & r$_{av}$ &I                 &r$_a$/r$_s$&i \\
No.   &            & M$_J$& R$_J$ &AU &   &days      &  UA      &10$^6$erg/cm$^{2}$s&           &deg \\
\hline

 124&HD4208b&  0.8 & - &  1.67&  0.05 &  812.197&1.6721&0.38 &10.6& - \\
 125&HD114386b&1.24& -&  1.65&  0.23 &  937.7  &1.6936&0.13  & -  & - \\
 126&HD45350b &0.98& -&  1.77&  0.78 &  890.76 &2.3084&0.35  &13.9& - \\
 127&HD222404b&1.7 & -&  2.13&  0.12 &  905    &2.1453& -    &   -& - \\
 128&HD213240b&4.5 & -&  2.03&  0.45 &  951    &2.2355&0.98  &23.1& - \\
 129&HD10647b &0.91& -&  2.1 &  0.18 &  1040   &2.1340&0.45  &33.2& - \\
 130&HD10697b &6.35& -&  2.12&  0.12 &  1072.3 &2.1353&1.02  &11.5& - \\
 131&HD95128b &2.54& -&  2.09&  0.06 &  1089   &2.0938&0.34  &3.5 & -  \\
 132&HD95128c &0.76& -&  3.73&  0.1  &  2594   &3.7486&0.11  &24.9& - \\
 133&HD190228b&3.58& -&  2.02&  0.499&  1146   &2.2715&0.95  &-   & - \\
 134&HD114729b&0.84& -&  2.08&  0.32 &  1135   &2.1865&0.62  &18.5& - \\
 135&HD111232b& 6.8& -&  1.97&  0.2  &  1143   &2.0094&0.23  &13.4& - \\
 136&HD2039b&   5.1& -&  2.2 &  0.69 &  1190   &2.7237&0.32  & -  & - \\
 137&HD136118&12.08& -&  2.4 &  0.36 &  1208.724&2.5555&0.61 &29.3& - \\
 138&HD50554b& 5.16& -&  2.41&  0.501&  1293   &2.7125&0.27  &28.0& -  \\
 139&HD196050b&3.02& -&  2.43&  0.3  &  1321   &2.5393&0.39  &24.6& - \\
 140&HD216437b&1.82& -&  2.32&  0.29 &  1256   &2.4175&0.52  &16.8& - \\
 141&HD216435b&1.49& -&  2.7 &  0.34 &  1442.919&2.8561&0.71 & -  & -  \\
 142&HD106252b&7.56& -&  2.7 &  0.471&  1600    &2.9995&0.19 &25.1& - \\
 143&HD23596b&  8.1& -&  2.88&  0.292&  1565    &3.0028&0.43 & -  & - \\
 144&HD145675b&4.74& -&  2.8 &  0.338&  1796.4  &2.9599&0.11 &16.7& - \\
 145&HD142022b&4.4 & -&  2.8 &  0.57 &  1923    &3.2549&0.13 &19.9& - \\
 146&HD39091b &10.3& -&  3.28&  0.61 &  2049    &3.8902&0.12 & -  & - \\
 147&HD72659b &2.55& -&  3.24&  0.18 &  2185    &3.2925&0.31 &26.7& - \\
 148&HD70642b &2   & -&  3.3 &  0.1  &  2231    &3.3165&0.17 & -  & - \\
 149&HD33636b&10.58& -&  4.08&  0.55 &  2928    &4.6971&0.07 &52.8&  - \\
 150&HD22049b&0.92 & -&  3.4 &  0.43 &  2548.667&3.7143&0.05 &51.1& -  \\
 151&HD117207b&2.06& -&  3.78&  0.16 &  2627    &3.8284&0.16 &20.3& - \\
 152&HD30177b&9.17 & -&  3.86&  0.3  &  2819.654&4.0337&0.1  & -  & - \\
 153&HD50499b&1.71 & -&  3.86&  0.23 &  2482.7  &3.9621&0.17 &29.7& -  \\
 154&HD190360b&1.502& -& 3.92&  0.36 &  2891    &4.1740&0.09 &23.6& -  \\
 155&HD190360c&0.057& -& 0.128& 0.01 &  17.1    &0.1280&101.7&0.57& - \\
 156&HD89307b &2.73 & -& 4.149& 0.27 & 3090     &4.3002&0.16 &-   &-  \\

\hline\hline
       \end{tabular}
  \end{center}
\end{table*}
\linespread{1.}

\
\end{document}